\documentclass[11pt]{article}
\usepackage[margin=2 cm]{geometry}
\usepackage{mathtools}
\usepackage{slashed}
\usepackage{graphicx}

\title{Lorentz violation in $\gamma$-pair production ($e^{+}e^{-} \rightarrow \gamma\gamma$)}
\author{Alessio T. B. Celeste and Adriano M. Santos\\Instituto Federal do Sertão Pernambucano, Campus Serra Talhada\\56915-899, Rodovia PE 320, Km 126, Serra Talhada-PE, Brazil}
\date{}
 
\begin{document}

\maketitle

\begin{abstract}
In this paper, we investigate the production process of a photon pair in a regime of Quantum Electrodynamics (QED), considering the presence of a constant four-vector that defines a privileged direction in spacetime, resulting in Lorentz symmetry violation and CPT violation. We explore Lorentz violation in the context of vectorial nonminimal coupling, where we analyze the modifications introduced by this coupling in the scattering amplitude at the vertex. Moreover, we delve into Lorentz violation with axial-like nonminimal coupling. Both analyses are conducted by examining the constant four-vector of both time-like and space-like types.
\end{abstract}
%\noindent{\it Keywords\/}: photon-pair production; constant four-vector; Lorentz violation.\\

\section{Introduction}
Symmetry is a concept widely investigated in multiple collections of human knowledge \cite{Hon}. In Physics, the concept of symmetry naturally leads us to the concept of invariance by some kind of transformation \cite{Noether,Wigner}. A notable example is the Lorentz symmetry, which holds great prominence among them all, playing a key role in Einstein's relativity \cite{Einstein}. Lorentz symmetry is a fundamental principle of special relativity, which states that the laws of physics are the same in any inertial reference frame. However, at energies approaching the Planck scale $M_{Planck} \sim 10^{19}$ GeV, where Quantum Gravity becomes important, Lorentz symmetry may be violated \cite{Brown}. This is an intriguing feature that has motivated the search for possible Lorentz symmetry violations, which is an active area of research in high-energy physics \cite{Ted}, cosmology \cite{Chico}, astrophysics \cite{Desai} and many other areas. Gaining a comprehensive understanding of the fundamental aspects of the universe under extreme energy conditions, while considering the implications of quantum gravity \cite{Moffat}, stands as a pivotal pursuit within the scope of this scientific investigation. Furthermore, the Planck scale is considered the smallest possible scale at which current physics can describe nature \cite{Igor}, making the study of Lorentz symmetry violation one of the most challenging and exciting areas of physics. 

In recent decades, the search for new physical theories has grown considerably, which inevitably results in the extension of the Standard Model with the breaking of Lorentz symmetry \cite{Carroll}. The beginning of this search for new formulations in terms of symmetry, in physical theories, intensifies with the construction of the Standard-Model Extension (SME) by Kostelecký and Colladay \cite{Colladay1,Colladay2} and with it the possibility of collapsing the Lorentz covariance in the context of the Quantum Field Theory that has been extensively studied since then. As a result of these studies from the late 1980s onwards, Lorentz symmetry breaking became a topic present in several situations, from supersymmetric models \cite{Berger,Carlson} to higher-derivative models in field theory \cite{Gomes, Mariz, Gomes2} passing through systems in condensed matter \cite{Belich1,Belich2,Charneski,Alan} and so on. 
In this work, we will dedicate attention to one of these contexts within the scope of scattering processes in Quantum Electrodynamics (QED) \cite{Colladay3,Altschul}. 

Experimental investigations aimed at verifying the validity of Lorentz-violating (LV) corrections have been extensively carried out following the advancements in theoretical studies, leading to the establishment of multiple constraints on LV parameters \cite{Iranianos}. Among these experiments, the clock anisotropy stands out as one of the most precise spectroscopic techniques \cite{Brown}, enabling the derivation of upper bounds of $10^{-33}$ GeV on LV parameters when incorporated within the framework of the SME. Nevertheless, studies on the possible effects of Lorentz Invariance Violation (LIV) on scattering processes are scarce, particularly regarding the determination of upper bounds on the breaking parameters through cross-sectional analyses. As regards the scattering process of producing $\gamma$-pairs, which is one of the fundamental reactions in QED, it has been extensively examined in collider experiments (see \cite{Derrick}). The validation of theoretical predictions necessitates the substantiation through empirical evidence. While acknowledging the inherent challenges in promptly securing such evidence, the endeavors outlined in the  tables \cite{Kostelecky1} provide a preliminary glimpse into potential avenues for validation. 

This paper is organized as follows. In section 2, a brief analysis of the $\gamma$-pair production scattering in the context of QED is conducted by considering contributions from Feynman diagrams at the tree-level. The electroweak interaction describes this scenario, and the objective is to express the differential cross section to lowest-order in terms of the scattering angle. In section 3, the investigation explores Lorentz violation with vectorial nonminimal coupling. The gauge-invariant vectorial nonminimal coupling involves the dual electromagnetic tensor $\tilde{F}_{\mu\nu}$, a coupling constant $g$, and a constant four-vector $b_{\mu}$ with dimensions of inverse mass.  The scattering amplitude calculation is modified to incorporate a new vertex, leading to the Lorentz-violated cross section expressed in terms of Mandelstam variables. Lastly, in section 4, a different result is obtained by considering Lorentz violation with axial-like nonminimal coupling. 

\section{Photon-pair production scattering}
\begin{center}
\begin{figure}[!ht]
\centering
\includegraphics[scale=.5]{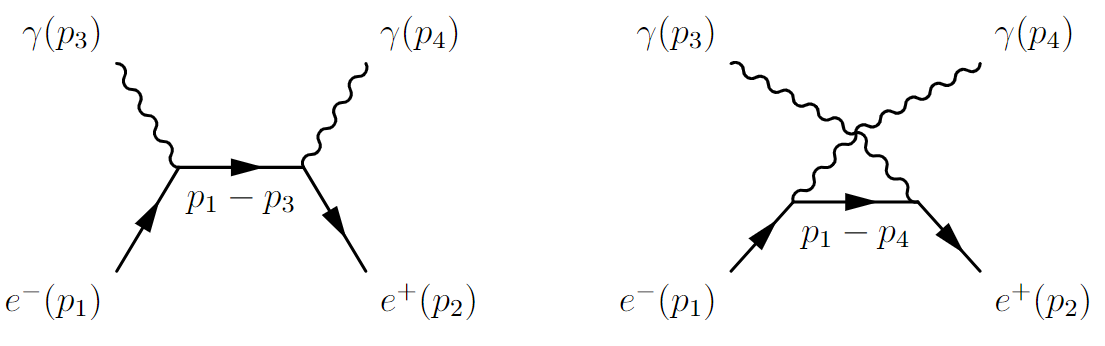}
\caption{Scattering $e^{+}e^{-} \rightarrow \gamma\gamma$ (left) t-channel, (right) u-channel.}
\label{figure1}
\end{figure}
\end{center}

A brief approach in QED of the $\gamma$-pair production scattering will be analyzed in this section. The scattering will be examined from the two contributions tree-level of the Feynman diagrams (t-channel and u-channel) displayed in Fig. \ref{figure1}. At higher energies, the contribution of heavier particles may appear beyond the diagrams, this scenario is described by the electroweak interaction.\\

These diagrams contribute to the calculation of the scattering amplitude. The momentum vectors are chosen so that $p_{1,2} = (E, 0, 0, \pm p), \, p_{3,4} = (E, \pm E \sin\theta, 0, \pm E \cos\theta)$\footnote{Azimuth angle $\phi$ cancels out in scattering calculations.}, in center of mass frame, $E=\sqrt{p^2 + m^2}$ is total energy and $m$ is electron (positron) mass ($\sim$ 0,5 MeV).

Using $u(p)$ for the electron spinor, $\bar{v}(p)=v^{\dagger}(p)\gamma^{0}$ for adjoint spinor positron  and $\epsilon^{*}(p)$ to denote the polarization vectors of the photons (outgoing), we have for the S-matrix element in Figure \ref{figure1}:
\begin{eqnarray}
i\mathcal{M}_{0} &=& i\mathcal{M}_{0}^{t} + i\mathcal{M}_{0}^{u} \nonumber\\
&=& \bar{v}(p_2)(-ie\gamma^\mu)\epsilon^{*}_{\mu}(p_4)\frac{i(\slashed{p}_1 - \slashed{p}_3 + m)}{(p_1 - p_3)^2 - m^2} \epsilon^{*}_{\nu}(p_3)(-ie\gamma^\nu)u(p_1) + \nonumber\\  
&& \bar{v}(p_2)(-ie\gamma^\nu)\epsilon^{*}_{\nu}(p_3)\frac{i(\slashed{p}_1 - \slashed{p}_4 + m)}{(p_1 - p_4)^2 - m^2} \epsilon^{*}_{\mu}(p_4)(-ie\gamma^\mu)u(p_1),
\end{eqnarray}
%}
%
where $e$ is the charge of the electron, $-ie\gamma^\mu$ is vertex factor usual and the free-fermion propagator is given by
\begin{eqnarray}
S(q)= \frac{i}{\slashed{q} - m} = \frac{i(\slashed{q} + m)}{q^2 - m^2},
\label{fermions_propagator}
\end{eqnarray}
which represents the continuous inner line in Feynman diagrams. In the regime of QED ($< 100$ MeV) and in the ultrarelativistic limit $E = E_{cm}/2 \cong |\mathbf{p}| \gg m$, we still consider that $p^2 = m^2 = 0$. Therefore, the unpolarized differential cross section, that is, without regard to spin and polarization orientations can be written in the form\footnote{The symbols $s, t$ and $u$ are the Mandelstam variables: $s = (p_1 + p_2)^2 = (2E)^2 = E_{cm}^{2}, t = (p_1 - p_3)^2 = 2E_{cm}^{2} (cos\theta -1), u = (p_1 - p_4)^2 = - 2E_{cm}^{2} (cos\theta + 1)$, where $E_{cm}$ is the energy of the center of mass, $\theta$ is the scattering angle.}
\begin{eqnarray}
\left(\frac{d \sigma}{d \Omega}\right)_{QED} = \frac{1}{64 \pi^2 E_{cm}^2} \left\langle |\mathcal{M}|^2 \right\rangle = \frac{\alpha^2}{2 s} \left( \frac{u}{t} + \frac{t}{u} \right),
\end{eqnarray}
where $\alpha = e^2/4\pi \approx 1/137$ is fine-structure constant and natural units were used. We can leave still the differential cross section in terms of the scattering angle, so we can also write to the lowest-order differential cross section for $\gamma$-pair production
\begin{eqnarray}
\label{secaochoque}
\frac{d \sigma}{d\,cos\theta} = \frac{2 \pi \alpha^2}{s} \left( \frac{1 + \cos^2\theta}{\sin^2\theta} \right).
\end{eqnarray}

\section{Lorentz violation with vectorial nonminimal coupling}
The Lorentz-violating Lagrange density of interest in this section is given by \cite{Belich1}
\begin{eqnarray}
\label{vectorial}
\mathcal{L}_{LIV} &&= \mathcal{L}_{QED} + \mathcal{L}_{vect} \nonumber\\
     &&= \bar{\psi}\left(i \gamma^{\mu}\partial_{\mu} - m\right)\psi - \frac{1}{4} F_{\mu\nu}^{2} - e \bar{\psi} \gamma^{\mu} \psi A_{\mu} + g b^{\nu}\bar{\psi}\gamma^{\mu}\psi \Tilde{F}_{\mu\nu}.
\end{eqnarray}
Note that the last term represents an invariant gauge operator with mass dimension 5 that violates Lorentz and CPT, this vectorial nonminimal couplings represents a subset of the operator with coefficient $a_{F}^{(5)\mu\alpha\beta}$ of the SME \cite{Kostelecky1}. $\Tilde{F}^{\mu\nu} = \frac{1}{2} \epsilon^{\mu\nu\alpha\beta}F_{\alpha\beta}$ is the dual electromagnetic tensor and the field strength is defined by $F_{\alpha\beta} = \partial_{\alpha}A_{\beta} - \partial_{\beta}A_{\alpha}$ with $\epsilon^{0123} = 1$ (Levi-Civita symbol), $g$ a coupling constant and $b^{\mu}$ a constant four-vector with dimensions of inverse mass. This four-vector defines a privileged direction in space-time which causes the violation of Lorentz symmetry and CPT. The presence of this last term in the Lagrangian generates a modification in the vertices as follows
\begin{eqnarray}
\label{newvertice}
i\Gamma^{\mu} = i e \gamma^{\mu} - g b^{\chi} \gamma^{\nu} q^{\alpha} \epsilon_{\chi\nu\alpha\beta} \eta^{\beta\mu},
\end{eqnarray}
where $q$ representing the four-momentum carried by the fermion line.

\begin{center}
\begin{figure}[!ht]
\centering
\includegraphics[scale=.5]{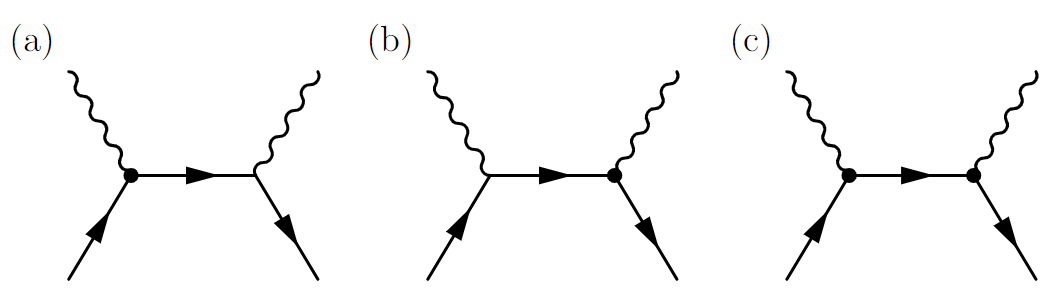}
\caption{Lorentz violation $e^{+}e^{-} \rightarrow \gamma\gamma$ in t-channel. In (a) and (b) one vertex with LV, (c) two vertices with LV.} 
\label{figure2}
\end{figure}
\end{center}

The diagrams of the (a) and (b) of the Fig. \ref{figure2} for t-channel will contribute to the scattering amplitude with
\begin{eqnarray}
i\mathcal{M}^{t}_{a} = \bar{v}(p_2)\left(i g b^{\chi} \gamma^{\tau}(p_1 - p_3)^{\alpha}\epsilon_{\chi\tau\alpha\beta}\eta^{\beta\mu}\right)\epsilon^{*}_{\mu}(p_4)\frac{i(\slashed{p}_1 - \slashed{p}_3 + m)}{(p_1 - p_3)^2 - m^2}\epsilon^{*}_{\nu}(p_3)(-ie\gamma^\nu)u(p_1),
\end{eqnarray}
and
\begin{eqnarray}
i\mathcal{M}^{t}_{b} = \bar{v}(p_2)(-ie\gamma^\mu)\epsilon^{*}_{\mu}(p_4)\frac{i(\slashed{p}_1 - \slashed{p}_3 + m)}{(p_1 - p_3)^2 - m^2}\epsilon^{*}_{\nu}(p_3)\left(i g b^{\chi} \gamma^{\tau}(p_1 - p_3)^{\alpha}\epsilon_{\chi\tau\alpha\beta}\eta^{\beta\nu}\right)u(p_1).
\end{eqnarray}

Note that diagrams (a) and (b) contribute equally to the calculation of the scattering amplitude. For diagram (c) in the Fig. \ref{figure2}, we have
\begin{eqnarray}
i\mathcal{M}^{t}_{c} &=& \bar{v}(p_2)\left(i g b^{\chi} \gamma^{\tau}(p_1 - p_3)^{\alpha}\epsilon_{\chi\tau\alpha\beta}\eta^{\beta\mu}\right)\epsilon^{*}_{\mu}(p_4) \nonumber\\
&\times& \frac{i(\slashed{p}_1 - \slashed{p}_3 + m)}{(p_1 - p_3)^2 - m^2} \epsilon^{*}_{\nu}(p_3) \left(i g b^{\chi} \gamma^{\delta}(p_1 - p_3)^{\kappa}\epsilon_{\chi\delta\kappa\sigma}\eta^{\sigma\nu}\right) u(p_1).
\end{eqnarray}

\begin{center}
\begin{figure}[!ht]
\centering
\includegraphics[scale=.5]{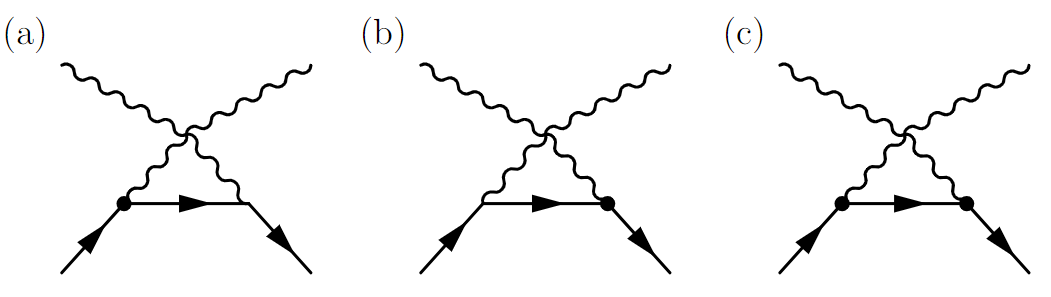}
\caption{Lorentz violation $e^{+}e^{-} \rightarrow \gamma\gamma$ in u-channel. In (a) and (b) one vertex with LV, (c) two vertices with LV.} 
\label{figure3}
\end{figure}
\end{center}

The diagrams of the (a) and (b) of the Fig. \ref{figure3} for u-channel will contribute to the scattering amplitude with
\begin{eqnarray}
i\mathcal{M}^{u}_{a} = \bar{v}(p_2)\left(i g b^{\chi} \gamma^{\tau}(p_1 - p_4)^{\alpha}\epsilon_{\chi\tau\alpha\beta}\eta^{\beta\mu}\right)\epsilon^{*}_{\mu}(p_3)\frac{i(\slashed{p}_1 - \slashed{p}_4 + m)}{(p_1 - p_4)^2 - m^2}\epsilon^{*}_{\nu}(p_4)(-ie\gamma^\nu)u(p_1),
\end{eqnarray}
and
\begin{eqnarray}
i\mathcal{M}^{u}_{b} = \bar{v}(p_2)(-ie\gamma^\nu)\epsilon^{*}_{\nu}(p_3)\frac{i(\slashed{p}_1 - \slashed{p}_4 + m)}{(p_1 - p_4)^2 - m^2}\epsilon^{*}_{\mu}(p_4)\left(i g b^{\chi} \gamma^{\tau}(p_1 - p_4)^{\alpha}\epsilon_{\chi\tau\alpha\beta}\eta^{\beta\mu}\right)u(p_1).
\end{eqnarray}
In the same way that we observed in the t-channel, the diagrams (a) and (b) in the Fig. \ref{figure3} for u-channel contribute in the same way to the calculation of the scattering amplitude. For diagram (c) in the Figure \ref{figure3}, we have
\begin{eqnarray}
\label{sechoqvl}
i\mathcal{M}^{u}_{c} &=& \bar{v}(p_2)\left(i g b^{\chi} \gamma^{\tau}(p_1 - p_4)^{\alpha}\epsilon_{\chi\tau\alpha\beta}\eta^{\beta\mu}\right)\epsilon^{*}_{\mu}(p_3) \nonumber\\
&\times& \frac{i(\slashed{p}_1 - \slashed{p}_4 + m)}{(p_1 - p_4)^2 - m^2}\epsilon^{*}_{\nu}(p_4) \left(i g b^{\chi} \gamma^{\delta}(p_1 - p_4)^{\kappa}\epsilon_{\chi\delta\kappa\sigma}\eta^{\sigma\nu}\right)u(p_1).
\end{eqnarray}

Making the necessary modifications in the calculation of the scattering amplitude with the replacement of this new vertex, Eq. (\ref{newvertice}), the Lorentz-violated cross section written in terms Mandelstam variables it becomes Eq. (\ref{sechoqvl2}).
\begin{eqnarray}
\left(\frac{d \sigma}{d \Omega}\right)_{LIV} = \left(\frac{d \sigma}{d \Omega}\right)_{QED} + \left(\frac{d \sigma}{d \Omega}\right)_{vect},
\label{sechoqvl2aa}
\end{eqnarray}
where the second term is being shown in the Eq. (\ref{sechoqvl2aa})

%{\fontsize{10}{10}\selectfont
\begin{eqnarray}
\left(\frac{d \sigma}{d \Omega}\right)_{vect} = \frac{\alpha^2}{4s} \Bigg\{\frac{u}{t} (b \cdot p_1 - b \cdot p_3)^3(b \cdot p_2 - b \cdot p_1) - b^4 (t^2 - 3tu + u^2) + \nonumber\\
t b^2 \Big[3 (b \cdot p_1)^2 + 3 (b \cdot p_3)^2 + (b \cdot p_4)^2 - (b \cdot p_1)(3 b \cdot p_2 + 4 b \cdot p_3 + 3b \cdot p_4) + \nonumber\\
(b \cdot p_2) (b \cdot p_4 - 2 b \cdot p_3)\Big] - 2 \Big[(b \cdot p_1)^4 + (b \cdot p_3)^4 + (b \cdot p_4)^4\Big] + \nonumber\\
u b^2 \Big[3 (b \cdot p_1)^2 + (b \cdot p_3)^2 + 3 (b \cdot p_4)^2 - (b \cdot p_1) (3 b \cdot p_2 + 3 b \cdot p_3 + 4 b \cdot p_4) + \nonumber\\
(b \cdot p_2) (b \cdot p_3 - 2\cdot p_4)\Big] + (b \cdot p_1)^3 (6 b \cdot p_2 + 5 b \cdot p_3 + 5 b \cdot p_4) - \nonumber\\
(b \cdot p_1)^2\Big[7 (b \cdot p_3)^2 + 7 (b \cdot p_4)^2 + 4 (b \cdot p_3)(b \cdot p_4) + 5 (b \cdot p_2) (b \cdot p_3 + b \cdot p_4)\Big] + \nonumber\\
(b \cdot p_1)\Big[(b \cdot p_4)^2(b \cdot p_3 - b \cdot p_2) - (b \cdot p_3)^2(b \cdot p_2 - 6 b \cdot p_3) + 6 (b \cdot p_4)^3  + \nonumber\\
(b \cdot p_3)(b \cdot p_4)(4b \cdot p_2 + b \cdot p_3) \Big] + \frac{t}{u} (b \cdot p_1 - b \cdot p_4)^3(b \cdot p_2 - b \cdot p_1) + \nonumber\\
(b \cdot p_2) (b \cdot p_3 + b \cdot p_4) \Big[2 (b \cdot p_3)^2 - 3 (b \cdot p_3)(b \cdot p_4) - 2 (b \cdot p_4)^2\Big]\Bigg\}. 
\label{sechoqvl2}
\end{eqnarray}
%}

In the case of a purely time-like background, i.e, $b_\mu = (b_0, \vec{0})$, the Eq. (\ref{sechoqvl2}) will be
{\fontsize{11}{11}\selectfont
\begin{eqnarray}
\left(\frac{d \sigma}{d \Omega}\right)^{t}_{LIV} = \left(\frac{d \sigma}{d \Omega}\right)_{QED} &-& e^4 (g b_0)^4 \Big[4E^2(t+u) + t^2 + u^2 +tu\Big] - 2e^4(g b_0)^2 \Big(8E^2 + t + u\Big).
\label{tipotempovetorial}
\end{eqnarray}
}
We emphasize that in models that use the lagrangian (\ref{vectorial}) with the electromagnetic tensor instead of its dual, the purely time-like background does not generate contributions for LIV. In the case of a purely space-like background, i.e, $b_\mu = (0, \vec{b})$, the Eq. (\ref{sechoqvl2}) will be

\begin{eqnarray}
\left(\frac{d \sigma}{d \Omega}\right)^{s}_{LIV} = \left(\frac{d \sigma}{d \Omega}\right)_{QED} + e^4 g^2 \Big\{2 (t+u) b^2 -4 (b \cdot p_3)^2 - 4(b \cdot p_4)^2 + \frac{g^2}{t u} \Big[-(t+u)^2 (b \cdot p_1)^4 + \nonumber\\
(b \cdot p_1)^3 \Big(u (b \cdot p_3) (5t + 3u) + (b \cdot p_2)(t^2 + 6ut + u^2) + (b \cdot p_4)(3t^2 + 5tu) \Big) - \nonumber\\
(b \cdot p_1)^2 \Big[(b \cdot p_4)^2(3t^2 + 7tu) + (b \cdot p_4)\Big(4 (b \cdot p_3)tu + (3t^2 + 5tu) (b \cdot p_2)\Big) + \nonumber\\
(b \cdot p_3)(b \cdot p_2)(5tu + 3u^2) + (b \cdot p_3)^2(7tu + 3u^2)\Big] + \nonumber\\
(b \cdot p_1)\Big[(b \cdot p_4)^3(t^2 + 6tu) + (b \cdot p_4)^2\Big((b \cdot p_2)(3t^2 - tu) + tu(b \cdot p_3)\Big) + \nonumber\\
tu (b \cdot p_3)(b \cdot p_4)(4b \cdot p_2 + b \cdot p_3) + (b \cdot p_3)^2\Big((b \cdot p_3)(6tu + u^2) - \nonumber\\
(b \cdot p_2)(tu -3u^2)\Big)\Big] - 2tu(b \cdot p_4)^4 - (t^2 - 2tu)(b \cdot p_2)(b \cdot p_4)^3 - \nonumber\\
tu (b \cdot p_2)(b \cdot p_3)(b \cdot p_4)^2 - tu b^4 (t^2 + ut + u^2) + u (b \cdot p_3)^3 \Big((2t - u) (b \cdot p_2) - \nonumber\\
2t(b \cdot p_3)\Big) - tu(b \cdot p_2)(b \cdot p_3)^2 (b \cdot p_4) + tu b^2 \Big(3 (t+u) (b \cdot p_1)^2 - \nonumber\\
(b \cdot p_1)\Big(3(b \cdot  p_2)(t + u) + (4t + 3u) (b \cdot p_3) + (3t + 4u)(b \cdot p_4)\Big) + \nonumber\\
(3t + u)(b \cdot p_3)^2 + (t + 3u)(b \cdot p_4)^2 + (b \cdot p_2)\Big((b \cdot p_3)(u - 2t) + (t - 2u)(b \cdot p_4)\Big)\Big)\Big] + \nonumber\\
4(b \cdot p_2)(b \cdot p_3 + b \cdot p_4) + 4(b \cdot p_1)(2b \cdot p_2 + b \cdot p_3 + b \cdot p_4)\Big\}.
\end{eqnarray}

Note that when $s \to \infty$ this result preserves unitarity, unlike the result of time-like four-vector found in Eq (\ref{tipotempovetorial}).

\section{Lorentz violation with axial-like nonminimal coupling}

The Lorentz-violating Lagrange density of interest in this section is given by
\begin{eqnarray}
\label{axial}
\mathcal{L}_{LIV} &=& \mathcal{L}_{QED} + \mathcal{L}_{axial} \nonumber\\ &=& \bar{\psi}\left(i \gamma^{\mu}\partial_{\mu} - m\right)\psi - \frac{1}{4} F_{\mu\nu}^{2} - e \bar{\psi} \gamma^{\mu} \psi A_{\mu} + g_5 b^{\nu}\bar{\psi}\gamma^{\mu} \gamma^{5} \psi \Tilde{F}_{\mu\nu},
\end{eqnarray}
where the term $\mathcal{L}_{axial}$ represents a axial-like nonminimal coupling, due to the presence of the matrix $\gamma^5$. The diagrams of the (a) and (b) of the Figure \ref{figure3} for u-channel will contribute to the scattering amplitude with

{\fontsize{11}{11}\selectfont
\begin{eqnarray}
i\mathcal{M}^{t}_{a,axial} = \bar{v}(p_2)\left(i g_5 b^{\tau} \gamma^{\nu} \gamma^5 (p_1 - p_3)^{\alpha}\epsilon_{\tau\nu\alpha\beta}\eta^{\beta\mu}\right) \epsilon^{*}_{\mu}(p_4)\frac{i(\slashed{p}_1 - \slashed{p}_3 + m)}{(p_1 - p_3)^2 - m^2}\epsilon^{*}_{\nu}(p_3)(-ie\gamma^\nu)u(p_1), 
\end{eqnarray}
}
and
{\fontsize{11}{11}\selectfont
\begin{eqnarray}
i\mathcal{M}^{t}_{b,axial} = \bar{v}(p_2)(-ie\gamma^\mu)\epsilon^{*}_{\mu}(p_4)\frac{i(\slashed{p}_1 - \slashed{p}_3 + m)}{(p_1 - p_3)^2 - m^2}\epsilon^{*}_{\nu}(p_3) \left(i g_5 b^{\tau} \gamma^{\zeta} \gamma^5 (p_1 - p_3)^{\alpha}\epsilon_{\tau\zeta\alpha\beta}\eta^{\beta\nu}\right)u(p_1).
\end{eqnarray}
}
For diagram (c) in the Figure \ref{figure3}, we have
{\fontsize{11}{11}\selectfont
\begin{eqnarray}
i\mathcal{M}^{t}_{c,axial} &=& \bar{v}(p_2)\left(i g_5 b^{\tau} \gamma^{\zeta} \gamma^5 (p_1 - p_3)^{\alpha}\epsilon_{\tau\zeta\alpha\beta}\eta^{\beta\mu}\right)\epsilon^{*}_{\mu}(p_4) \nonumber\\
&\times& \frac{i(\slashed{p}_1 - \slashed{p}_3 + m)}{(p_1 - p_3)^2 - m^2} \epsilon^{*}_{\nu}(p_3)\left(i g_5 b^{\omega} \gamma^{\delta} \gamma^5 (p_1 - p_3)^{\kappa}\epsilon_{\omega\delta\kappa\sigma}\eta^{\sigma\nu}\right) u(p_1).
\end{eqnarray}
}

The diagrams of the (a) and (b) of the Figure \ref{figure3} for u-channel will contribute to the scattering amplitude wit
{\fontsize{11}{11}\selectfont
\begin{eqnarray}
i\mathcal{M}^{u}_{a,axial} = \bar{v}(p_2)\left(i g_5 b^{\tau} \gamma^{\zeta} \gamma^5 (p_1 - p_4)^{\alpha}\epsilon_{\tau\zeta\alpha\beta}\eta^{\beta\mu}\right)\epsilon^{*}_{\mu}(p_3) \frac{i(\slashed{p}_1 - \slashed{p}_4 + m)}{(p_1 - p_4)^2 - m^2}\epsilon^{*}_{\nu}(p_4)(-ie\gamma^\nu)u(p_1),
\end{eqnarray}
}
and
{\fontsize{11}{11}\selectfont
\begin{eqnarray}
i\mathcal{M}^{u}_{b,axial} = \bar{v}(p_2)(-ie\gamma^\nu)\epsilon^{*}_{\nu}(p_3)\frac{i(\slashed{p}_1 - \slashed{p}_4 + m)}{(p_1 - p_4)^2 - m^2}\epsilon^{*}_{\mu}(p_4) \left(i g_5 b^{\tau} \gamma^{\zeta} \gamma^5 (p_1 - p_4)^{\alpha}\epsilon_{\tau\zeta\alpha\beta}\eta^{\beta\mu}\right)u(p_1).
\end{eqnarray}
}
For diagram (c) in the Figure \ref{figure3}, we have
{\fontsize{11}{11}\selectfont
\begin{eqnarray}
i\mathcal{M}^{u}_{c,axial} &=& \bar{v}(p_2)\left(i g_5 b^{\tau} \gamma^{\zeta} \gamma^5 (p_1 - p_4)^{\alpha}\epsilon_{\tau\zeta\alpha\beta}\eta^{\beta\mu}\right)\epsilon^{*}_{\mu}(p_3) \nonumber\\
&\times& \frac{i(\slashed{p}_1 - \slashed{p}_4 + m)}{(p_1 - p_4)^2 - m^2}\epsilon^{*}_{\nu}(p_4)\left(i g_5 b^{\omega} \gamma^{\delta} \gamma^5 (p_1 - p_4)^{\kappa}\epsilon_{\omega\delta\kappa\sigma}\eta^{\sigma\nu}\right)u(p_1).
\end{eqnarray}
}
Thus, the Lorentz-violated cross section written in terms Mandelstam variables it become
\begin{eqnarray}
\label{sechoq5vl}
\left(\frac{d \sigma}{d \Omega}\right)_{LIV} = \left(\frac{d \sigma}{d \Omega}\right)_{QED} + \left(\frac{d \sigma}{d \Omega}\right)_{axial},
\end{eqnarray}
where the second term is being shown in the Eq. (\ref{sechoq5vl2}).
{\fontsize{10}{10}\selectfont
\begin{eqnarray}
\left(\frac{d \sigma}{d \Omega}\right)_{axial} = \frac{\alpha^2 g_5^4}{4s} \Bigg\{\frac{u}{t} (b \cdot p_1 - b \cdot p_3)^3(b \cdot p_2 - \,\, b \cdot p_1) +  \frac{t}{u} (b \cdot p_1 - b \cdot p_4)^3(b \cdot p_2 - b \cdot p_1) + \nonumber\\
t b^2 \Big[ 3 (b \cdot p_1)^2 + 3 (b \cdot p_3)^2 + (b \cdot p_4)^2 - (b \cdot p_1)(3 b \cdot p_2 + 4 b \cdot p_3 + 3 b \cdot p_4) + \nonumber\\
(b \cdot p_2) (b \cdot p_4 - 2 b \cdot p_3) \Big] -2 \Big((b \cdot p_1)^4 + (b \cdot p_3)^4 + (b \cdot p_4)^4\Big) + 2b^2(t + u) +\nonumber\\
u b^2 \Big[3 (b \cdot p_1)^2 + (b \cdot p_3)^2 + 3 (b \cdot p_4)^2 - (b \cdot p_1)(3b \cdot p_2 + 3b \cdot p_3 + 4 b \cdot p_4) + \nonumber\\
(b \cdot p_2) (b \cdot p_3 - 2 b \cdot p_4) \Big] + (b \cdot p_3)^2 \Big((b \cdot p_1) (b \cdot p_2)-4\Big) + 5 (b \cdot p_1)^3 (b \cdot p_3 + b \cdot p_4) +\nonumber\\
b \cdot p_2 \Big[ 6(b \cdot p_1)^3 - 5 (b \cdot p_1)^2(b \cdot p_3 + b \cdot p_4) - (b \cdot p_1)\Big((b \cdot p_3)^2 - 4(b \cdot p_4) (b \cdot p_3) + (b \cdot p_4)^2 - 8\Big) + \nonumber\\
(b \cdot p_3 + b \cdot p_4) \Big(2 (b \cdot p_3)^2 - 3 b \cdot p_4 (b \cdot p_3) + 2 (b \cdot p_4)^2 + 4\Big) \Big] - b^4(t^2 + t u + u^2) - \nonumber\\
(b \cdot p_1)^2 \Big[7 (b \cdot p_3)^2 + 4 (b \cdot p_4) (b \cdot p_3) + 7 (b \cdot p_4)^2\Big] - 4 (b \cdot p_2) (b \cdot p_3) (b \cdot p_4) (b \cdot p_1) + \nonumber\\
b \cdot p_1 \Big[(b \cdot p_3 + b \cdot p_4) \Big(6 (b \cdot p_3)^2 - 5 (b \cdot p_4) (b \cdot p_3) + 6 (b \cdot p_4)^2 + 4\Big) + (b \cdot p_2) \Big(4 (b \cdot p_4) (b \cdot p_3) -\nonumber\\
(b \cdot p_3)^2 - (b \cdot p_4)^2 + 8\Big)\Big] + (b \cdot p_4)^2 \Big((b \cdot p_2) (b \cdot p_1) -7 (b \cdot p_1)^2 -4\Big) - 8 (b \cdot p_2) (b \cdot p_1) \Bigg\}.
\label{sechoq5vl2}
\end{eqnarray}
}

In the case of a purely time-like background, i.e, $b_\mu = (b_0, \vec{0})$, the Eq. (\ref{sechoq5vl2}) will be
{\fontsize{11}{11}\selectfont
\begin{eqnarray}
\left(\frac{d \sigma}{d \Omega}\right)^{t}_{LIV} = \left(\frac{d \sigma}{d \Omega}\right)_{QED} - \,\, (e g_5 b_0)^4 \Big[ 4 E^2 (t+u) + t^2 + tu + u^2\Big]. 
\label{tipotempoaxial}
\end{eqnarray}
}
Note that in the same way to the time-like case the above result contains only even terms in $(g_5b_0)$. In the case of a purely space-like background, the Eq. (\ref{sechoq5vl2}) will be

{\fontsize{10}{10}\selectfont
\begin{eqnarray}
\left(\frac{d \sigma}{d \Omega}\right)^{s}_{LIV} = \left(\frac{d \sigma}{d \Omega}\right)_{QED} 
-\frac{e^4 g_5^4}{t u} \Big\{u \Big[(b \cdot p_2) \Big((u-2 t) (b \cdot p_3)^3+8 (t^3+t)\Big)-8 t^3 (b \cdot p_3)+2 t (b \cdot p_3)^4\Big]-\nonumber\\
(b \cdot p_1)^3 \Big[(t^2+6 t u+u^2) (b \cdot p_2)+u (5 t+3 u) (b \cdot p_3)+t (3 t+5 u) (b \cdot p_4)\Big]+(t+u)^2 (b \cdot p_1)^4+\nonumber\\
(b \cdot p_1)^2 \Big[t (3 t+7 u) (b \cdot p_4)^2+t (b \cdot p_4) \Big((3 t+5 u) (b \cdot p_2)+4 u (b \cdot p_3)\Big)+\nonumber\\
u (b \cdot p_3) \Big((5 t+3 u) (b \cdot p_2)+(7 t+3 u) (b \cdot p_3)\Big)\Big]+(b \cdot p_1) \Big[t (b \cdot p_4)^2 \Big((u-3 t) (b \cdot p_2)-u (b \cdot p_3)\Big) -\nonumber\\
t (t+6 u) (b \cdot p_4)^3 - t u (b \cdot p_3) (4 b \cdot p_2+b \cdot p_3) (b \cdot p_4)+u (b \cdot p_3)^2 \Big((t-3 u) (b \cdot p_2)-\nonumber\\
(6 t+u) (b \cdot p_3)\Big)\Big]+2 t u (b \cdot p_4)^4+t (t-2 u) (b \cdot p_2) (b \cdot p_4)^3+\nonumber\\
t u (b \cdot p_2) (b \cdot p_3) (b \cdot p_4)^2 + t u (b \cdot p_4) \Big((b \cdot p_2) (b \cdot p_3)^2-8\Big)+t u b^2 \Big[-3 (t+u) (b \cdot p_1)^2+\nonumber\\
(b \cdot p_1) \Big(3 (t+u) (b \cdot p_2)+(4 t+3 u) (b \cdot p_3)+(3 t+4 u) (b \cdot p_4)\Big)-(3 t+u) (b \cdot p_3)^2-\nonumber\\
(t+3 u) (b \cdot p_4)^2+(b \cdot p_2) \Big((2 t-u) (b \cdot p_3)-(t-2 u) (b \cdot p_4)\Big)\Big] + t u (b^4) (t^2+t u+u^2)\Big\}.
\end{eqnarray}
}

As in the previous section, note that this result preserves unitarity, unlike the case of time-type four-vector found in Eq. (\ref{tipotempoaxial}).

\section{Conclusion}
In this paper, we have investigated Lorentz symmetry breaking by analyzing the  $\gamma$-pair production scattering in the scenario of Extended Quantum Electrodynamics. Through our analysis, we have uncovered valuable insights into the nature of coupling and its implications. Specifically, in the context of vectorial coupling, our examination of the results of cross-sections revealed the emergence of fourth-order and second-order terms in the coupling constant for the time-like and space-like fields, respectively. Similarly, in the case of axial coupling, our findings demonstrated the presence of exclusively fourth-order terms in the coupling constant for the time-like field, while the space-like field exhibited second-order terms. Notably, one intriguing observation from our study is the consistent appearance of a fourth-order magnitude in the coupling constant for the axial time-like scenario, similar to the behavior observed in the standard case. This contrast to the fourth and second-order magnitudes observed in the coupling constants of other scenarios presents a potentially intriguing phenomenon. Finally, we hope that these results may be useful as a guide in the investigation of Lorentz symmetry breaking. The emergence of fourth-order and second-order terms in the coupling constants for different scenarios, as revealed in our analysis of  $\gamma$-pair production scattering, provides valuable insights for further research.

\end{document}